# AN EFFECTIVE APPROACH TO REDUCE THE PENETRATION POTENTIAL OF SARS-COV-2 AND OTHER VIRUSES BY SPIKE PROTEIN: SURFACE PARTICLE ELECTROSTATIC CHARGE NEGOTIATION

Kausik Rakshit[a], Sudip Chatterjee[a*], Durjoy Bandyopadhyay[a], Somsekhar Sarkar[b]

a) Swami Vivekananda University, Regent Education and Research Foundation, Barrack Pore, Kolkata 700121, India

b) 2nd Year, Integrated BS-MS UG student, Indian Institute of Science Education and Research, KOLKATA

*corresponding author's e mail : sudip4734@gmail.com

sudipc@regent.ac.in

**Abstract:** The objective of this paper is to provide a mathematical model to construct a barrier that may be useful to prevent the penetration of different viruses (Eg. SARS-COV-2) as well as charged aerosols through the concept of electrostatic charge negotiation. (Fusion for the opposite types of charges and repulsion for the similar types of charges).

Reviewing the works of different authors, regarding charges, surface charge densities ($\sigma$), charge mobility ($\mu$) and electrostatic potentials of different aerosols under varied experimental conditions, a similar intensive study has also been carried out to investigate the electron donating and accepting (hole donating) properties of the spike proteins (S-proteins) of different RNA and DNA viruses, including SARS-COV-2.

Based upon the above transport properties of electrons of different particles having different dimensions, a mathematical model has been established to find out the penetration potential of those particles under different electrostatic fields.

An intensive study have been carried out to find out the generation of electrostatic charges due to the surface emission of electrons (SEE), when a conducting material like silk, nylon or wool makes a friction with the

Gr IV elements like Germanium or Silicon, it creates an opposite layer of charges in the outer conducting surface and the inner semiconducting surface separated by a dielectric materials. This opposite charge barriers may be considered as Inversion layers (IL). The electrostatic charges accumulated in the layers between the Gr IV Ge is sufficient enough to either fuse or repel the charges of the spike proteins of the RNA, DNA viruses including COVID-19 (RNA virus) or the aerosols.

1) **Theoretical Background, Journal study and Investigation:**

**1.1 Calculation of Charges of Different Aerosols and Graphical Analysis:**

A study by **M.V.Rodrigues[1] et al** on the values of charges accumulated on different aerosol particles found, that the accumulation of the charges and charge densities($\sigma$) largely depend upon the Stokes diameter of the particle. The variation of charges with the Stokes diameter has been found most likely to be following a linear graph with a definite slope. The results have been used extensively in our study.

However, it has also been observed that for a larger particle having diameter more than d > 4 μm, abruptly different levels of charge density are present.

In the experimental setup consisting of an ECC (Schematics in adjoining figure), a set of values for the charge accumulation on dry egg dust particles and their respective Stokes radius was found. It is referred below in **Table 1**.

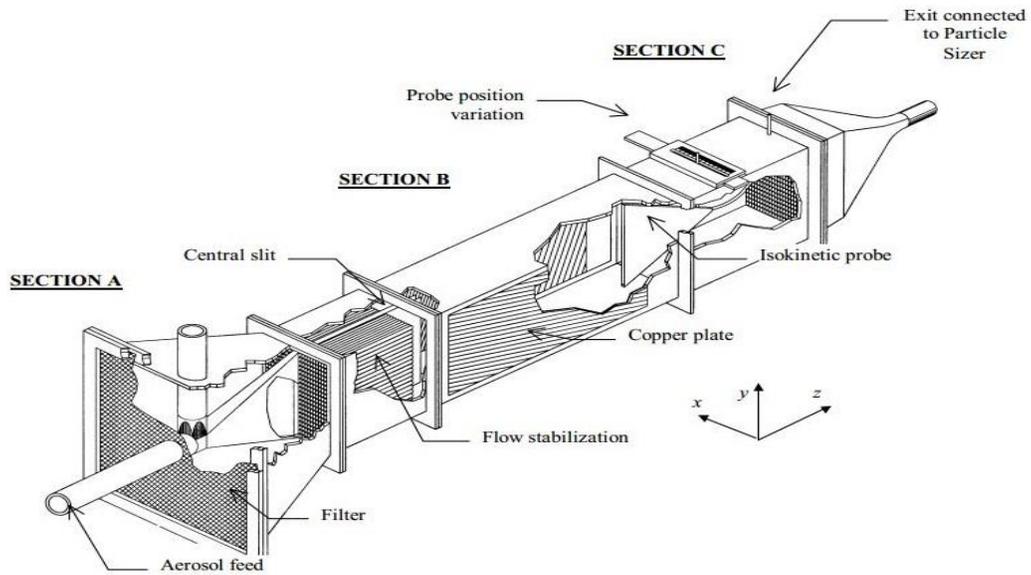

General view of the Electrical Charge Classifier (ECC).

Table 1

| Stokes Diameter [$10^{-6}$m] | Charge [$10^{-17}$C] |
|---|---|
| 1.7 | 0.5 |
| 2.1 | 0.5 |
| 2.5 | 1.2 |
| 3.2 | 1.6 |
| 4.8 | 3.0 |

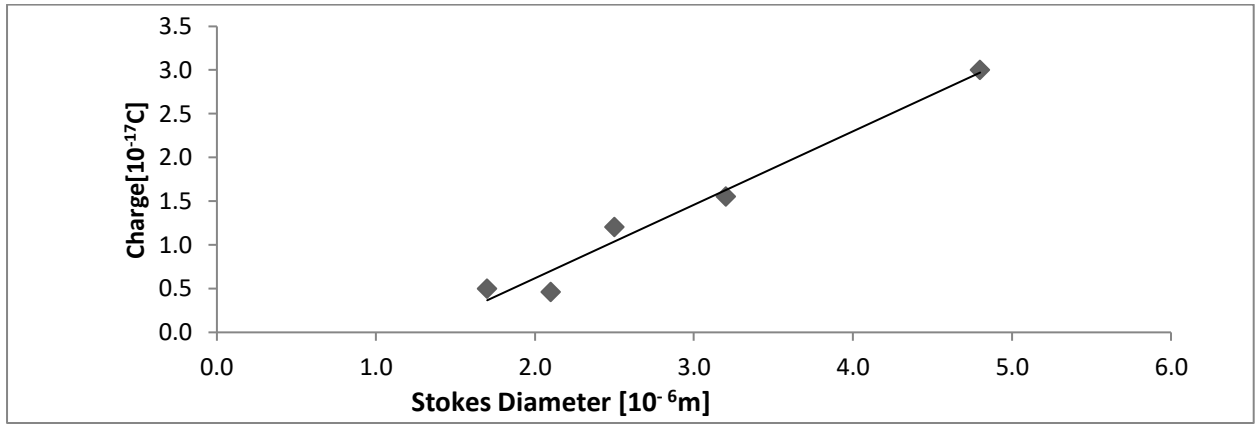

*Figure I: Particle charge as a function of Stokes diameter for aerosol (Concentrated dry egg dust particle - Data set in Table 1)*

The data from **Table 1** is fitted to the equation 1.1.1 (by Jonhston et al. 1987):

**|Q/e|=Ad$_p^B$**　　　　　　　　　　　　　　**[1.1.1]**

Where **e** is the elementary charge of an electron and **A** is the median number of elementary charges of magnitude **e** present on a particle of diameter 1 μm., It is found that, the value of **B** is nearly equal to 2 and also at per as suggested by Jonhston. (Range between 1 and 2).

This expected range of **B** is also applicable for calculating the charge of any small particles.

**1.2 Theoretical investigation for the factors, influencing the charge on SARS-COV-2 surfaces:**

The accumulation of charges on the surface of **SARS-COV-2** can be attributed by the following four factors:

- **The size and shape of the virus:** Literature study [**Angeletti et al[4]**] suggests, **SARS-COV-2** has a spherical or ellipsoidal shape having average diameter in the range of 60 nm to 140 nm. This makes a prediction that the pattern of charge on the surface of **SARS-COV-2** is similar to that of aerosols as established in **1.1**.

- **The (pKa)s of the protein on the surface:** At the isoelectric point of a protein, there is an equivalent distribution of negative and positive charges, leading to a neutral response to a potential difference. However, if the pH be lowered from the isoelectric point, it is found that, a bias is created towards the centre of positive charge of the protein. The pKa's could be derived from the pH values of the titration curve for the S(spike)-protein of the SARS-CoV-2 using the Henderson–Hasselbalch equation (from titration analysis):

    **pH = pK$_a$ + log ([conjugate base]/[weak acid])**           [1.2.1]

To establish this concept of the influence of surrounding pH on the surface charge of a virus, similar experiments had already been carried out also over the other virus strains.

**Relevant examples:**

Poliovirus 1 (strain LSc) was established to be isoelectric in nature at pH level of 6.6. **[Zerda et al[5]]**. A second strain of poliovirus 1 (strain Brunhilde) had been found isoelectric at the level of pH 7.1. Phage MS-2 was isoelectric at pH 3.9. Also, reovirus type 3 has been reported to be isoelectric at pH 3.9 by column focusing techniques **[Zerda et al[5]]**. Hence, it can be concluded that, there is a tendency of viruses to accumulate different charges on the surface area at different pH levels.

- **The protein size of the outermost spikes of the virus:** Also considering that the 1255 long amino acid sequence of the S-protein has been found and the corresponding possible oxidizing state of the contributing amino acids can be visualized, one may actually determine exactly how much charge is accumulated on each protein subunit at a particular pH level.

As stated earlier, actual fractional dissociation ($\alpha_i$) of any group which can be ionized is related to pH by the Henderson−Hasselbalch equation:

$$\mathbf{pH = pK_i + \log\left[\frac{\alpha i}{1-\alpha i}\right]} \qquad [1.2.2]$$

Where, $pK_i$ may be calculated if $\alpha_i$ at corresponding pHs are known. The $pK_i$ depicts the negative logarithm of the effective dissociation constant. In the usual charge determination procedure **[Jakubke and Jeschkeit, 1977; Skoog and Wichman, 1986]**., the p$Ki$ for each type of ionizable group is assigned a magnitude. The verified knowledge of the amino acid sequence of a protein is used to calculate the net charge, $z_P$, as -$(z_P)_i$, where the charge of protein arising from $n_i$ ionizable groups of type $i$, $(z_p)i$, is given by:

$$\mathbf{(z_P)_i = n_i\, z_i\, \alpha_i} \qquad [1.2.3a]$$

where the ionizable group, $i$, is anionic ($z_i$ is negative)

and, $\qquad\qquad\qquad \mathbf{(z_P)_i = n_i\, z_i\, (1 - \alpha_i)} \qquad\qquad [1.2.3b]$

where, the ionizable group, $i$, is cationic (since, $z_i$ is positive).

However, p$Ki$ cannot be assigned a fixed magnitude because of its dependence on the overall charges of the protein. It is more difficult to dissociate a proton from a negatively charged molecule than from one with a net positive charge. This variation of p$Ki$ can be taken into account **[Compton and O'Grady, 1991]** by means of the theoretical expression for proton dissociation that has been in existence for over 80 years **[Linderstrøm-Lang 1924; Linderstrøm-Lang and Nielsen, 1959]**. Specifically, p*Ki* can be expressed by:

**$pK_i = (pK_{int})_I - 0.868\ wz_p$**  [1.2.4]

The Spike protein (S Protein) is a large type of transmenbrane protein ranging from 1160 amino acid for Avian infectious Bronchitis virus (IBV) and upto 1400 amino acids for Feline Corona virus (FCo) [**Taylor Heald-Sargent, et al[21]**]. It also has been revealed from different studies of **Donald J.Winzor[7] and Harold P. Erickson[8]** that, approximate numbers of electrons can effectively be calculated from the ratio of the mass of the spike proteins and that of the amino acids since it may be considered that each of the amino acid can carry maximum a single donating or accepting charge. (depending upon their shapes and sizes). The ultimate structural understanding of a protein comes from an atomic-level structure obtained by X-ray crystallography or nuclear magnetic resonance (NMR) spectroscopy. However, structural information at the nanometer level is frequently invaluable. Hydrodynamics, in particular sedimentation and gel filtration, can provide this structural information, and it becomes even more powerful when combined with electron microscopy (EM).

A hypothetical visualization is that if we consider one charged site per 'n' amino acids at a particular pH, the number of free electrons present per nano meter length of spike proteins remain in the order of (1400/n) and the value of the charges per nano meter length of the spike may be of the order of $1.6 \times 10^{-19} \times (1400/n) = (2240 \times 10^{-19}/n)$ C. (Experimental determination of the value of n is beyond the scope of this paper and is strongly recommended by the Authors). Therefore, at a particular pH

**Charge per nm length of spike protein = $(2240 \times 10^{-19}/n)$ C**  [1.2.5]

Considering the above values, a mathematical model has been established in **1.3** to find out the charges of the spike proteins of different lengths ranging from 8.0 nm to 10.0 nm for SARS-CoV-

2. It has been revealed by **[Zerda et al[5]]**. Therefore, it can be predicted that, the spike proteins of SARS-CoV-2 shows measurable electron donating or accepting capabilities.

- **Maintenance of Membrane Potential across Enveloping protein:**

If we look at the enveloping proteins of the SARS-CoV, refereeing to the following research by *Dewald Schoeman et al[45].*, we have found that, the Envelope Proteins (E-proteins) have a capability to maintain a neutrality through the membrane potential of their own, till a considerable amount of charge difference is created when the structure of whole envelope collapses.

The COV E protein is a short, integral membrane protein of 76–109 amino acids, ranging from 8.4 to 12.0 kDa in size. The primary and secondary structure reveals that, E protein has a short, hydrophilic amino terminus consisting of 7–12 amino acids, followed by a large hydrophobic Trans Membrane Domain (TMD) of 25 amino acids, and ends with a long, hydrophilic carboxyl terminus, which comprises the majority of the protein. The hydrophobic region of the TMD contains at least one predicted amphipathic α-helix that oligomerizes to form an ion-conductive pore in membranes.

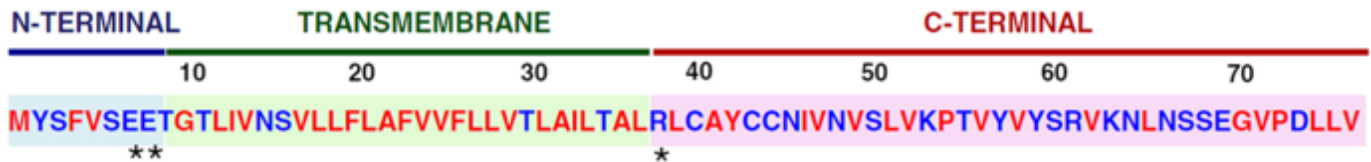

**Amino Acid Sequence and Domains of the SARS-CoV E Protein. Amino acid properties are indicated: hydrophobic (red), hydrophilic (blue), polar, charged (asterisks)**

Comparative and phylogenetic analysis of SARS-COV E revealed that, a substantial portion of the TMD consists of the two nonpolar, neutral amino acids, valine and leucine, lending a strong hydrophobicity to the E protein. The peptide exhibits an overall net charge of zero, the middle region being uncharged and flanked on one side by the negatively charged amino (N)-terminus, and, on the other side, the carboxy (C)-terminus of variable charge. The C-terminus also exhibits some hydrophobicity but less than the TMD due to the presence of a cluster of basic, positively charged amino acids.

An investigation regarding maintenance of this membrane potential had been done in case of Semiliki Forest Virus **[Andre V. Samsonov et al[6] ]** which is also an RNA virus, where it was established that, at a relatively positive charge, the potential barrier of the E1 and E2 envelope had been collapsed and the virus was spilled out by it's contents. Hence, it might be abstracted that coronavirus will also show similar results (Experimental determination of which is beyond scope of this theoretical work.) and that, there is a tendency among RNA viruses to develop (by any process) a general neutral Hydrophobic nature until the potential difference is very high, when the envelope system collapses.

**1.3 Calculation of the charges of the S-proteins of different coronaviruses viruses including SARS-COV-2 based on similarity to aerosols and their Graphical analysis:**

The virus responsible for COVID-19 is SARS-CoV-2 which belongs to the category of β corona virus. The structure of these viruses are found mostly spherical or ellipsoidal in nature having diameter in the range of 60nm to 140 nm, additionally having spikes in the range of length 8 nm to 10 nm. [Point to be noted: The aerosol particles referred to in 1.1 also have a diameter in the range of 50 nm to 100 nm.]

It should be considered here that the S-protein in the spikes of the coronavirus is essentially a complex folded structure of a chain formed from the the S gene that comprises an ORF encoding a protein of 1353 amino acid residues, with a predicted molecular weight of 149,918 *(S Mounir [1], P Labonté, P J Talbot)*. It tends to have different charge accumulations and configurations in different ph levels of the surroundings, unlike simple inorganic particles of aerosols.

Now, reaction to pH level is actually generated due to the difference of electric potential energy level of an electron from the energy level of the 1s orbital of the $H^+$. Similar possible different energy state of electron is obtained from bringing a molecule near to a charged surface. Hence this similarity has been exploited in this investigation. According to the recent research work[5,7], mutation in the spike protein is probably responsible for jumping this virus (after mutation) from species to human and [**Angeletti et al[4]**] proposed that the ORF1ab is the largest gene in SARS-CoV-2 which encodes the pp1ab protein and 15 nsps. This means that a structure forming such a major bulk of the virus is expected to have significant effects on the charge accumulation on the virus surface, behaving in closed spaces not unlikely to that in certain pH's.

The charges of the spike proteins have been measured in different laboratories by different methods but all of them have established a concrete correlation with the length and diameter of the spike proteins. However, it has been found that in most of the cases, the level of accumulation of charges show saturation beyond a certain length of the S-proteins. A possible explanation would be due to the destructive field interference of similar charges.

Presented below are some sets of data furnished from **equation [1.2.5]** with minimal standard deviation to extrapolate viable charge concentrations on coronaviruses:

**Accumulated charges per nm length of S-protein spikes on Coronavirus at a particular pH**

**(S-protein in the range of 8 nm to 10 nm)**

| Spike length | (Charge X n) |
|---|---|

| (nm) | (X $10^{-19}$ C) |
|---|---|
| 8.1 | 1.296 |
| 8.2 | 1.311 |
| 8.3 | 1.318 |
| 8.4 | 1.324 |
| 8.5 | 1.334 |
| 8.6 | 1.366 |
| 8.7 | 1.372 |
| 8.8 | 1.391 |
| 8.9 | 1.401 |
| 9.0 | 1.412 |
| 9.1 | 1.443 |
| 9.2 | 1.461 |
| 9.3 | 1.472 |
| 9.4 | 1.481 |
| 9.5 | 1.501 |
| 9.6 | 1.512 |
| 9.7 | 1.532 |
| 9.8 | 1.544 |
| 9.9 | 1.564 |
| 10.0 | 1.589 |

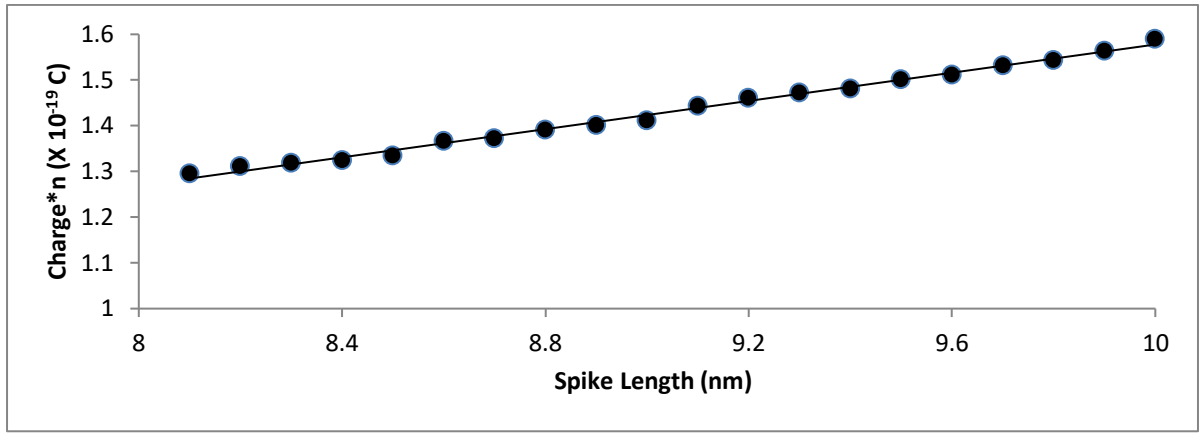

*Figure II: Particle charge as a function of Spike Length of SARS-COV-2*

Here, it can be seen that, the charge of the S-protein is predicted to vary linearly with spike length at a particular pH.

**2)      Mathematical model developed for calculating net charge on the surface of a single virus:**

**2.1 The factors affecting charge formation:**

The effects due to various factors affecting the charge are listed below in the form of equations stated beforehand:-

i.  Due to size of virus: (From 1.1.1)

$$|Q/e| = A d_p^B$$

ii. Due to interaction of S-protein with surrounding pH (from 1.2.3a):

$(z_P)_I = n_i \, z_i \, \alpha_i$

And from [1.2.3b]     $(z_P)_i = n_i \, z_i \, (1 - \alpha_i)$

From [1.2.2]     $pH = pK_i + \log[\frac{\alpha i}{1-\alpha i}]$

From [1.2.4]     $pK_i = (pK_{int})_I - 0.868 \, w z_p$

From the above four equations, we may derive a relation between the net charges on protein with the surrounding pH if $pK_i$ is known:

$$pH = [(pK_{int})_I - 0.868\, wz_p] + \log\left[\frac{(z_p)_i/n_i z_i}{1-((z_p)_i/n_i z_i)}\right] \quad \text{for anionic ionisable group} \quad [2.1.1a]$$

$$pH = [(pK_{int})_I - 0.868\, wz_p] - \log\left[\frac{(z_p)_i/n_i z_i}{1-((z_p)_i/n_i z_i)}\right] \quad \text{for cationic ionisable group} \quad [2.1.1b]$$

## 2.2 Derivation of probable formula to predict charge formation on virus:

At the isoelectric point of the S-protein of the virus, Equation 1.1.1 is the only factor affecting the charge formation. However, when the virus is in a different pH level from that of the isoelectric points of S-protein, Equation 2.1.1a and equation 2.1.1b are also applicable which affects the formation of charges. Hence at a particular pH, incorporating both the equations, the following mathematical model has been established.

From 1.1.1, $$Q = A e d_p^B$$

From 2.1, $$pH = [(pK_{int})_I - 0.868\, wz_p] \pm \log\left[\frac{(z_p)_i/n_i z_i}{1-((z_p)_i/n_i z_i)}\right]$$

$$\Rightarrow (pH - (pK_{int})_I + 0.868 wz_P) = \pm\log\left[\frac{(z_p)_i/n_i z_i}{1-((z_p)_i/n_i z_i)}\right]$$

$$\Rightarrow e^{\{\pm(pH-(pKint)I+0.868wzP)\}} = [((z_p)_i/n_i z_i)^{-1}] - 1$$

$$\Rightarrow (z_P)_i = (n_i z_i)[1 + e^{\{\pm(pH-(pKint)i+0.868wz_p)\}}]^{\wedge(-1)} \quad [2.2.1]$$

Here an assumption is made that, the charge formation on the S-protein due to pH change shall not affect the inherent tendency of the virus.

Thus, if the total charge at a particular pH on the virus be $Q_{tot}$, then from Equation 1.1.1 and Equation 2.2.1, it's value may be expressed as,

$$Q_{tot} = Q + [(z_p)_i]\{(\pi d_P^2)k\}$$

(Where k is the number of spikes per unit surface area of the virus. However, this value is not determined and the authors strongly suggest an experimental determination of this value

$Q_{tot} = Aed_p^B + [(n_i \cdot z_i)[1 + e^{\{\pm (pH-(pKint)I+0.868wzP\ )\}}]^{(-1)}] \{(\pi d_P^2)k\}$

$\Rightarrow Q_{tot} = (Ae)d_p^B + ([(n_i z_i)[1 + e^{\{\pm (pH-(pKint)I+0.868wzP\ )\}}]^{(-1)}]\{\pi k\})d_p^2$ [2.2.2]

Here it can be observed that, at a particular pH, the $Q_{tot}$ is still an exponential function of the diameter of the virus.

Now, if we relate the charge of the virus to the dry egg dust aerosol, it is expected to show similar types of properties as the virus is also behave like a dry egg dust particle, which is a rich in protein. Thus a value of B nearly equal to 2 can be expected here. Assuming B to be almost equal to 2, in this case the equation may be simplified for relatively smaller values of **$d_P$** as follows:

At a particular pH level,

**$Q_{tot} = (Ae + K) d_P^2$**

**Where, $K = [(n_i z_i)(1 + e^{\{\pm (pH-(pKint)I+0.868wzP\ )\}})^{(-1)}] (\pi k)$**

$\Rightarrow \boxed{Q_{tot} = \chi d_p^2}$ (2.2.3)

**Where, $\chi = Ae + [(n_i z_i)[1 + e^{\{\pm (pH-(pKint)I+0.868wzP\ )\}}]^{(-1)}] (\pi k)$**

This parameter '$\chi$' may be experimentally determined through suitable procedure. Hence, it would provide a fairly accurate prediction of the net charge on the surface of a virus at a particular pH level following the lines of this mathematical model.

Now, a particular potential difference may show very similar effect on ionizing proteins at a particular pH. Hence, this model may be extended for the calculation of charges on a virus surface under the influence of a particular potential gradient. In that case, let the parameter be 'C' (To be experimentally determined) and the equation be as follows:

$$Q_{tot} = Cd_p^2 \qquad (2.2.4)$$

**Fig III: This is a demonstration of nature of increase of $Q_{tot}$(depicted in vertical axis) with respect to $d_P$(depicted in horizontal axis)**

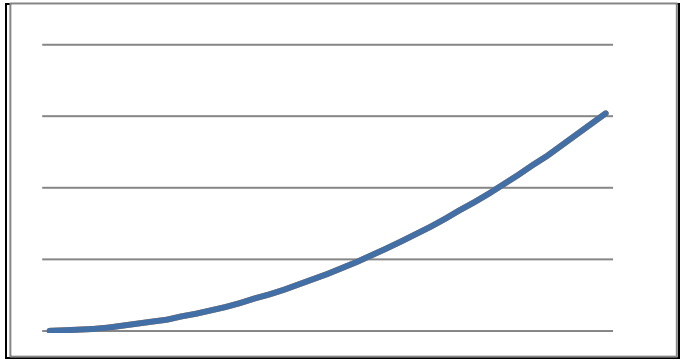

**3)   Prospective model of the potential barrier to stop viruses:**

**3.1 Formation of surface charge density (σ) over IL as a function of time:**

The electrostatic charges form due to the surface emission owing to the frictions between the layers of semiconducting materials like Ge with the conducting materials like nylon, wool.

When the conducting materials like wools make friction with a semiconducting Gr IV materials, it shows a surface emission of electrons and the charges remain accumulated at the outer conducting layers since the middle layer of the mask is a semiconductor and is separated by an effective dielectric medium.

The charge densities per square cm area(σ) follow empirically the growth of charges equation in a capacitor.

$$\sigma = [\varepsilon \mu N_f / 4\pi r^2] \left[\int (1 - e^{-t/CR}) dt\right] [(dS/dx)] \qquad [3.1.1]$$

Where, $\varepsilon$ is the permittivity of the medium, $\mu$ is the coefficient of friction between the two layers, $N_f$ is the normal force acting over the surface depending upon the rate of movement, $4\pi r^2$ is the surface area and dx is the separation distance between the conducting and semi conducting surface separated by the dielectric materials.

Given below is a graphical representation of **3.1.1**

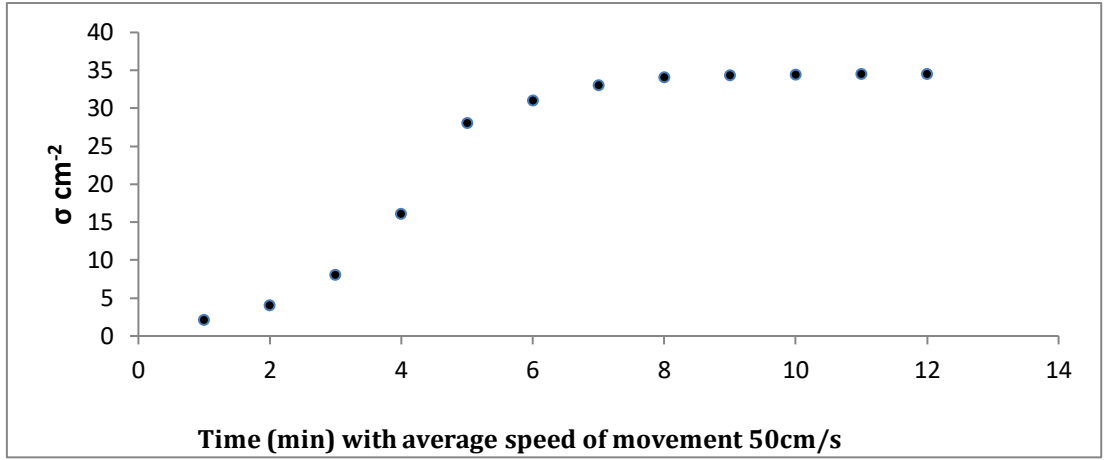

Fig III: Variation of accumulation of charge densities(σ) per cm$^{-2}$ with time in the IL due to average movement of 50 cm/s

Clearly, the density of accumulated charge is a time dependent, area dependent and movement dependent function. The produced charges will form an inversion layer (IL) at the outer and inner surface due to the induction. However the charge accumulation attains a finite quantity after a certain period of time, as can be seen from **Fig III**.

**3.2 Comparison of static charges produced in the layers of mask (IL) and the surface of the viruses and/or the air-borne aerosols:**

From the above studies it is clear that, the accumulation of charges per square cm area of a conducting bi layer due to the surface emission of electrons in normal movement is at least $10^6$ times more than the charges on surface of viruses including SARS-CoV-2 and also the charge density is much higher than the charges accumulated in the concentrated dust particles of normal ranges.

Therefore, it may be concluded that, due to the surface emission of electrons, the IL of a mask can effectively trap, neutralize or repel the the viruses and/or the charged aerosols which might be carrying air borne viruses.

4) **Conclusion:**

The charge accumulation on the surface of a virus is established to be in accordance with the **Equation 2.2.2**:

$$\boxed{Q_{tot} = (Ae)d_p^B + ([(n_i z_i)[1 + e^{\{\pm (pH-(pKint)I+0.868wzP\ )\}}]^{(-1)}]\ \{\pi k\})d_p^2}$$

Which has been simplified with some basic assumption to **Equation 2.2.3**:

$$\boxed{Q_{tot} = \chi d_p^2}$$

( Where, $\chi = [Ae + ([(n_i z_i)[1 + e^{\{\pm (pH-(pKint)I+0.868wzP\ )\}}]^{(-1)}]\ \{\pi k\})]$ )

Using a derivation from this mathematical model, the model of a protective barrier has been established as follows:

A three layer mask can be produced to prevent the immediate infections of DNA, RNA viruses including SARS-COV-2 and others because the electric charge accumulation of the RNA viruses is much less than the electrostatic charges accumulated in the layers of the mask within few minutes. In the inner surface, a cotton type non conducting material can be used which will work basically as a nonconductor so that the electrostatic charges produced inside the two layers does not drain out through surface of the (human)body. Additionally, the cotton layer may be effective to protect the skin from electrostatic thermal radiation.

Frictions between the middle and the outer layers, where the hydrophobic conducting and semiconducting materials are used, effectively lead to accumulation of the static charges.

The rate of accumulation of charges increase with the increase of friction but come to a saturation level following the model of growth of charges in a capacitor per square cm of area.

**5)  Suggestions for further experimental determinations:**

From the above studies, authors strongly recommend the following investigations:

- Experimental determination of the value of **'n'** as specified in **Equation 1.2.5**.
- Experimental determination of the value of **'k'** as specified in **Equation 2.2.2**.
- Experimental determination of parameters **'$\chi$'** and **'C'** as specified in this article.
- Conducting of further experiments to find out the applications of the same for the preparation of PPE or gloves.

**Acknowledgement:**

Authors are highly indebted to **Prof. (Dr.) Samir Kumar Patra**, Department of Life Science, National Institute of Technology Rourkela, and **Prof. (Dr.) Shantanu Bhowmik**, Dean (R&D) Aerospace Engineering Amrita University and Adjunct Professor TuDelft University, The Netherlands, for their constant encouragement and valuable suggestions.

*Corresponding author's email address: sudip4734@gmail.com*

*Corresponding author's email address: sudip4734@gmail.com*